\documentclass{JHEP3}
\preprint{LU-TP 04-41\\
  hep-ph/0412111}

\usepackage{cite}
\usepackage{epsfig}
\usepackage{xspace}
\usepackage{graphics}
\usepackage[latin1]{inputenc}

\def\hide#1{}

\newcommand{\pythia}{P\scalebox{0.8}{YTHIA}\xspace}

\newcommand{\smallx}{S\scalebox{0.8}{MALLX}\xspace}
\newcommand{\cascade}{C\scalebox{0.8}{ASCADE}\xspace}
\newcommand{\ldcmc}{\scalebox{0.8}{LDCMC}\xspace}

\newcommand{\kT}{\ensuremath{k_{\perp}}}
\newcommand{\pT}{\ensuremath{p_{\perp}}}

\newcommand{\kTi}[1]{\ensuremath{k_{\perp #1}}}

\newcommand{\gtaet}{\raisebox{-0.8mm}%
{\hspace{1mm}$\stackrel{>}{\sim}$\hspace{1mm}}}
\newcommand{\ltaeq}{\raisebox{-0.8mm}%
{\hspace{1mm}$\stackrel{<}{\sim}$\hspace{1mm}}}

\newcommand{\particle}[1]{\ensuremath{\mathrm{#1}}}
\newcommand{\antiparticle}[1]{\ensuremath{\bar{\mathrm{#1}}}}

\newcommand{\q}{\particle{q}}

\newcommand{\qbar}{\antiparticle{q}}

\newcommand{\qsq}{\ensuremath{Q^2}}

\def\mrm#1{\mathrm{#1}}

\def\sup#1{\ensuremath{^{\mrm{#1}}}}

\def\f2d3{\ensuremath{F_2^{\mrm{D}3}}}
\def\ftwo{\ensuremath{F_2}\xspace}
\def\done#1{}

\newcommand{\eqref}[1]{eq.~(\ref{#1})\xspace}
\newcommand{\eqsref}[1]{eqs.~(\ref{#1})\xspace}

\newcounter{aenumct}

{\end{list}}

\newcounter{enumct}

\newcommand{\leading}{\textit{leading}\xspace}
\newcommand{\gluonic}{\textit{gluonic}\xspace}
\newcommand{\standard}{\textit{standard}\xspace}

\skip\footins = 1\bigskipamount plus 2pt minus 4pt                              

\title{\boldmath Uncertainties on Central Exclusive Scalar
  Luminosities\\ from the unintegrated gluon distributions. }

\author{Leif Lönnblad and Malin Sjödahl\\
  Dept.~of Theoretical Physics,
  S\"olvegatan 14A, S-223 62  Lund, Sweden\\
  E-mail: \email{Leif.Lonnblad@thep.lu.se}
    and \email{Malin.Sjodahl@thep.lu.se}}
  
  \abstract{In a previous report we used the Linked Dipole Chain model
    unintegrated gluon densities to investigate the uncertainties in
    the predictions for central exclusive production of scalars at
    hadron colliders. Here we expand this investigation by also
    looking at other parameterizations of the unintegrated gluon
    density, and look in more detail on the behavior of these at small
    \kT. We confirm our conclusions that the luminosity function for
    central exclusive production is very sensitive to this behavior.
    However, we also conclude that the available densities based on
    the CCFM and LDC evolutions are not constrained enough to give
    reliable predictions even for inclusive Higgs production at the
    LHC.}

\begin{document}
 
\sloppy
 
\section{Introduction}
\label{sec:intro}

Detecting the Higgs boson at LHC in the ``most probable'' mass region
around 120 GeV is far from a trivial task, such a light Higgs
predominantly decays into bottom quarks making the background from
standard QCD processes huge. Looking for Higgs signals in the clean
environment of central diffractive events is therefore an appealing
prospect, provided the cross section is sufficiently
high\cite{Schafer:1990fz,Bialas:1991wj,Cudell:1996ki,
  Levin:1999qu,Khoze:2001xm,Cox:2001uq,
  Boonekamp:2002vg,Enberg:2002id}.

In general, central exclusive events can be used for studying any
scalar particle. In this paper we will only consider a Higgs boson,
but our results can be trivially generalized.  Central exclusive
diffractive Higgs production was first suggested in
\cite{Schafer:1990fz,Bialas:1991wj} and has lately been developed
further by Khoze, Martin and Ryskin (KhMR)\footnote{We shall here
  refer to their calculations as KhMR to distinguish it from the KMR
  procedure for obtaining unintegrated gluon densities from integrated
  ones by Kimber, Martin and
  Ryskin\cite{Kimber:2001sc}.}\cite{Khoze:2001xm}. One of the main
advantages compared to inclusive Higgs production is that, since the
central system is constrained to be in a $0^{++}$ state, the normal
QCD background from b-jets is heavily suppressed. By matching the mass
of the central system, as measured with the central detectors, with
the mass calculated from the energy loss of the scattered protons
detected by very forward proton taggers, it is possible to exclude
events with extra radiation outside the reach of the detectors, to
ensure that the central system is indeed in a $0^{++}$ state.

In \cite{Lonnblad:2003wx} we investigated the implications of the
uncertainties in the unintegrated structure functions, uPDFs, for the
KhMR calculations. Our main conclusion was that the cross section is
very sensitive to the unintegrated structure functions, $G(x,
k_\perp^2, m_H^2 )$, in the region of $k_\perp \approx 2-3$ GeV.  The
differences in the uPDF, which enters in the final exclusive
luminosity to the power of four, leeds to a variation in the result of
roughly one order of magnitude. This estimate was obtained using
unintegrated structure functions both from KMR\cite{Kimber:2001sc}
(used in the KhMR calculations) and different parameterizations based
on the Linked Dipole Chain model, LDC\cite{Gustafson:2002jy}.

In this report we have also used the CCFM-based densities described in
\cite{Hansson:2003xz}, here referred to as Jung-1 and Jung-2. Both LDC
and Jung have been tuned to \ftwo data from HERA in the region of
small $x\ltaeq 0.01$ and $1.5<Q^2\ltaeq 100$~GeV$^2$.  Despite the
similar fitting region and the similarities between CCFM and LDC
evolution, it is found that the densities differ substantially in
their \kT-distribution \cite{Gustafson:2002jy} even inside the fitting
region. This is to be expected, since the fitting was only done to
\ftwo, which is an integrated quantity. Below we will find that the
differences at high scales, corresponding to the production of a
$120$~GeV Higgs, is large even for the integrated density. This can
be explained by the fact that here the densities are also influenced
by the large x distribution at smaller scales, well outside the region
of the fit.

In the KMR case the uPDFs are derived directly from the globally
fitted integrated gluon density,
MRST98\cite{Martin:1998sq}\footnote{MRST98 is not the newest of PDF
  parameterizations, but it was used in \cite{Khoze:2001xm}, where it
  was also shown that the results are rather insensitive to the choice
  of integrated PDF.}. Hence at least the integrals of the uPDFs are
well constrained. On the other hand, the \kT\ dependence is uncertain
since KMR assumes DGLAP evolution which works well for inclusive
observables but not necessarily for \kT\ sensitive ones.

The off-diagonal unintegrated parton densities (oduPDFs) which enters
into the KhMR calcultaions were derived in \cite{Martin:2001ms} from
the corresponing off-diagonal integrated one (odPDF) in the same way
as the uPDFs were derived from the standard integrated PDFs in
\cite{Martin:1998sq}. Now, while the integrated gluon PDF is fairly
well constrained experimentally, the unintegrated is not, and the
off-diagonal unintegrated, used in the exclusive cross section, is
even less so. And any uncertainty in the uPDF will immediately be
reflected in an uncertainty in the oduPDF.


There are a few weak experimental constraints on the \kT-distribution
of the uPDFs. So far these constraints have not been taken into
account in any fitting, but comparing models using the uPDFs with data
can give us some hints about where the densities work and where they
need to be improved. Since the exclusive luminosity is sensitive to
the uPDF mainly in the region of a few GeV we should look for other
observables sensitive to features in this region to obtain
constraints.  One such observable is the \kT-spectra of W and Z in
hadron collisions (eg.\ at the Tevatron
\cite{Abbott:1999wk,Affolder:1999jh}) for small \kT. While the main
features of this can be reproduced by a calculation using KMR
uPDFs\cite{Watt:2003vf}, the small-\kT\ peak is slightly too low, as
can be seen in figure \ref{fig:watt}, indicating that KMR may be
underestimating somewhat the hardness of the \kT-distribution.

\FIGURE[t]{\includegraphics*[width=11cm,bb=275 191 475 321]{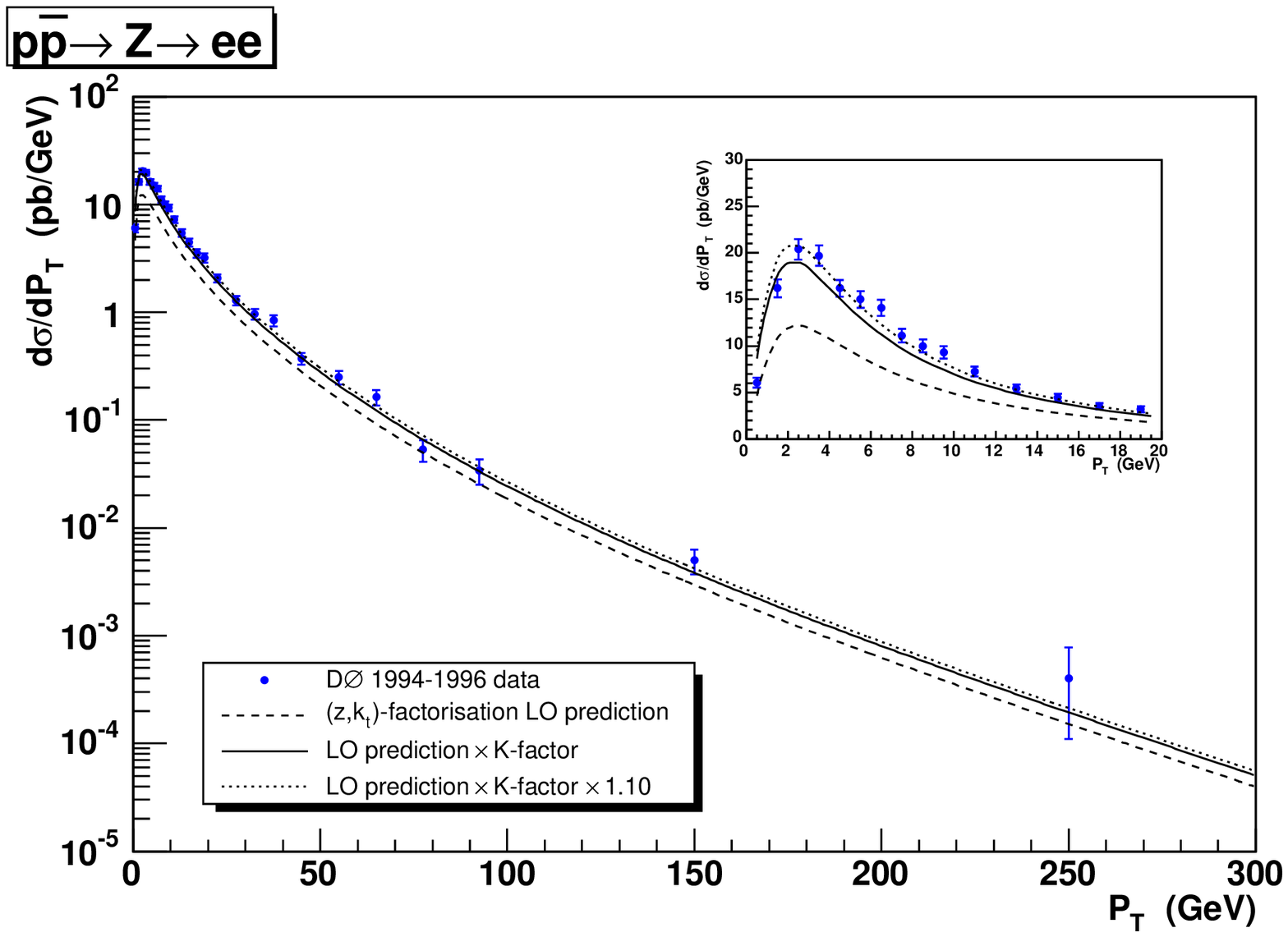}
  \caption{\label{fig:watt} The \pT-distribution of Z$^0$ measured at the
    Tevatron \cite{Abbott:1999wk} compared to a calculation using the
    KMR approach to uPDFs with different options as described in
    \cite{Watt:2003vf}.}}

Another sensitive observable is the rate of forward jets in DIS at
HERA. Especially in the measured region of $\kT^2\sim
Q^2\gtaet10$~GeV$^2$ and small $x$ where standard DGLAP evolution
would not contribute. Indeed, DGLAP based models severely
underestimate the rate of forward jets (see eg.\ 
\cite{Andersson:2002cf} and \cite{Andersen:2003xj} for a discussion on
this), and even though the KMR uPDFs have not been confronted with
this data it is likely that they will also fail.

In general there are indications of a slightly harder \kT\ distribution
in the uPDFs than what is given by KMR. This is predicted by the
BFKL-like CCFM evolution (and hence also LDC) on which the alternative
uPDFs used in this report are based on.  Such evolution includes also
ladders unordered in tranverse momenta, opening up for more activity.
As shown in \cite{Gustafson:2001iz} the typical evolution path,
starting from the high virtuality end, is a rapid DGLAP-like evolution
down to a few GeV and then a region of transverse momenta distributed
as a random walk in $\log(\kT)$.

The layout of this paper is as follows. First we recapitulate the main points 
in the calculation of Khoze, Martin and Ryskin and discuss their oduPDFs in 
section 2. In section 3 we obtain the oduPDFs in the case of LDC and Jung 
respectively. Then, in section 4 we present and comment our results. Finally 
we arrive at our conclusions in section 5.

\section{Central exclusive production}
\label{sec:exclusive}

\FIGURE[t]{%
  \begin{minipage}{11cm}
    \begin{center}
      \epsfig{file=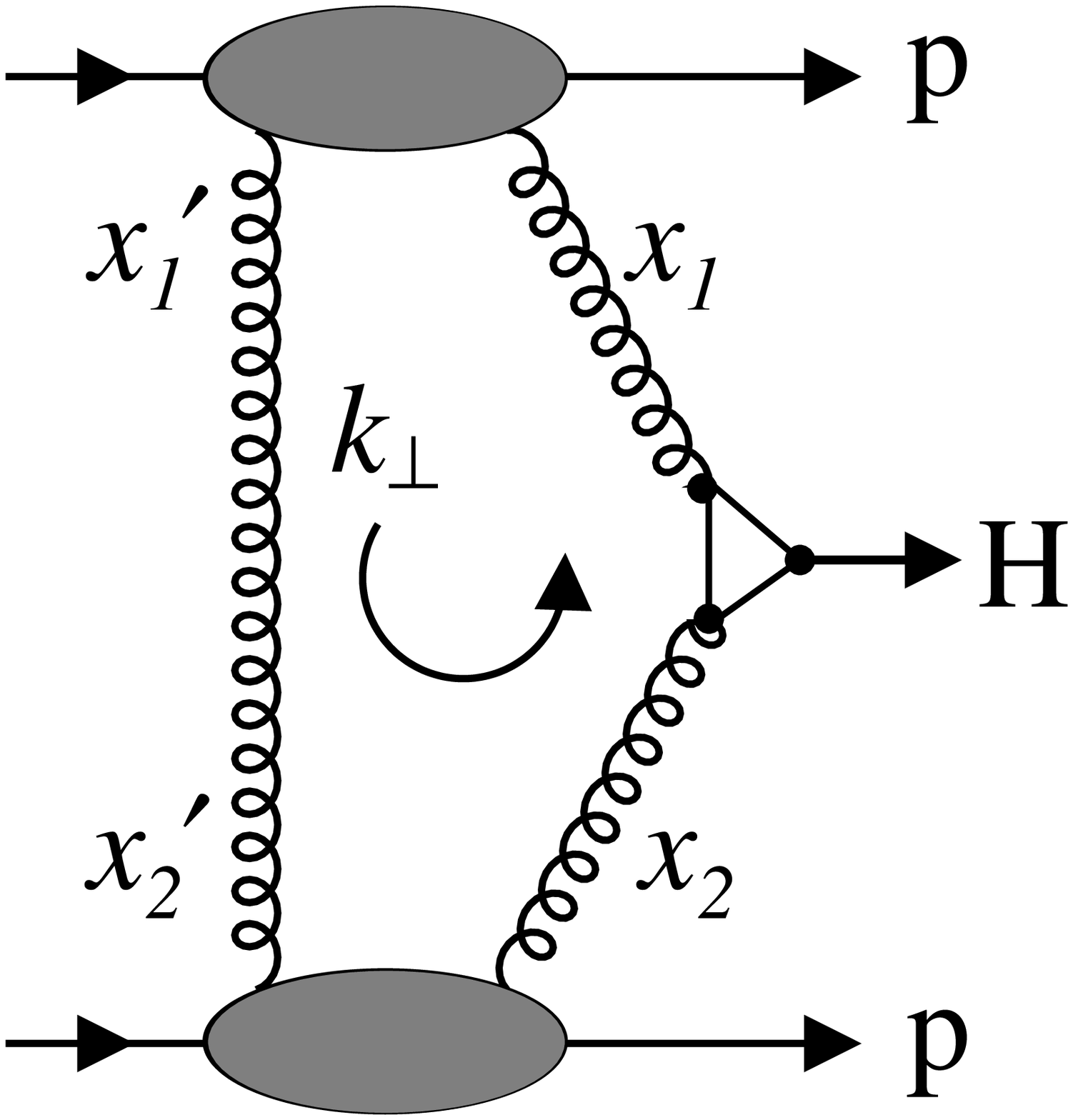,width=7cm}
    \end{center}
  \end{minipage}
  \caption{\label{fig:exdiagram} The basic diagram for exclusive production
    of the Higgs boson in hadron collisions.}}

In a central exclusive production of a Higgs boson, two gluons with no
net quantum number fuse into a Higgs via the standard heavy quark
triangle diagram, whereas another semi-hard gluon exchange guarantees
that there is no net colour flow between the protons. This is shown in
figure \ref{fig:exdiagram}, where it is also indicated that the
exchanged semi-hard gluon should compensate the transverse momentum
$k_\perp$ of the gluons producing the Higgs, so that the protons are
scattered with little or no transverse momenta.

Several types of radiation can destroy the diffractive character of
the interaction.  There can be extra interactions between the
spectator partons, modeled by the so called soft survival probability
$S^2$. Also, the gluons participating in the interaction can radiate
both at scales above \kT, which is modeled by the hard survival
probability using a Sudakov form factor, and at scales below \kT, which
is suppressed, since such gluons cannot resolve the the individual
colours of the exchanged gluon pair.

This is discussed in detail in \cite{Khoze:2002py} and
\cite{Lonnblad:2003wx}. Here we just state the resulting exclusive
luminosity function

\begin{eqnarray}
  \label{eq:kmrlum}
  L(M,y)&=&\frac{\delta^2 {\cal L}}{\delta y \delta \ln{M^2}}\\
  &=& S^2 \left[\frac{\pi}{(N_c^2-1)b}\int^{\mu^2} \frac{dk_\perp^2}{k_\perp^4}
    f_g(x_1,x_1',k_\perp^2,\mu^2)
    f_g(x_2,x_2',k_\perp^2,\mu^2)\right]^2\nonumber
\end{eqnarray}
where $\mu^2=M^2/4$ in the standard KhMR prescription, $y$ denotes
rapidity, $b$ comes from the probability for the protons to remain
intact, $x_{1(2)}=m_H e^{(-) y}$ and $x'_{1(2)}\sim \kT/\sqrt{S}\ll
x_{1(2)}$.  $f(x,x',k_\perp^2,\mu^2)$ is the off-diagonal unintegrated
gluon density, the oduPDF, which should be interpreted as the
amplitude related to the probability of finding two gluons in a proton
with equal but opposite transverse momentum, $k_\perp$, and carrying
energy fractions $x$ and $x'$ each, one of which is being probed by a
hard scale $\mu^2$.

The cross section is then obtained by

\begin{eqnarray*}
  \label{eq:sigma}
  \sigma = \int \hat{\sigma}_{gg\rightarrow H}(M^2)
  \frac{\delta^2 {\cal L}}{\delta y \delta\ln{M^2}}
  dyd\ln{M^2}
\end{eqnarray*}
where $M$ is the invariant mass of the central system, in this case
the Higgs mass.  In principle one should use a off-shell version of
$\hat{\sigma}$ (see eg.\ \cite{Hautmann:2002tu}) which then would have
a \kT\ dependence, and hence break the factorization. However, for the
exclusive cross section the main contribution comes from rather small
\kT\ and, at least for large masses, the factorization should hold.
Since the cross section, in the exclusive case, is a convolution of the
luminosity and the matrix element it suffices to study the difference
in luminosity to investigate the effects of different oduPDFs.

Besides the oduPDFs, the only other main uncertainty in
\eqref{eq:kmrlum} is the soft survival probability $S^2$. We have made
a separate study using the multiple interaction model in
\pythia\cite{Sjostrand:1987su,Sjostrand:2000wi} in the same way as was
done for the WW$\rightarrow$H process in \cite{Dokshitzer:1991he}.
Taking the probability of having no additional scatterings in Higgs
production\footnote{We here used inclusive Higgs production, but the
  result should be the same for the exclusive case} using the
parameters of the so-called Tune-A by Rick Field\cite{RickTuneA}, we
estimate the survival probability to be 0.040 for the Tevatron and
0.026 for the LHC. This is remarkably close to the values used in
\cite{Khoze:2001xm} obtained in the so-called two-channel eikonal
approach \cite{Kaidalov:2001iz}.

In the KhMR case the oduPDF \cite{Khoze:2001xm} are obtained in a two
step procedure presented in \cite{Martin:2001ms}.  In the first step
the off-diagonal parton distribution functions, odPDF, are extracted
from the standard gluon PDF, in the relevant limit of $x'\ll x$:
\begin{equation}
  \label{eq:pdf2odpdf}
  H(x,x',\mu^2)\approx R_g xg(x,\mu^2).
\end{equation}
Although we will use a constant $R_g$ factor of $1.2$, we note that it
in general depends on both $x$ and $\mu^2$.  The consequences for the
luminosity function of a non-constant $R_g$ are moderate and briefly
discussed in \cite{Lonnblad:2003wx}.

In the second step it is assumed that the oduPDF can be obtained from
the odPDF in the same way as the uPDF can be obtained from the
standard PDF. In the latter case one can use the KMR prescription
introduced in \cite{Kimber:2001sc}, where
\begin{equation}
  \label{eq:pdf2updf}
  G(x,k_\perp^2,\mu^2)\approx
  \frac{d}{d\ln k_\perp^2}\left[xg(x,k_\perp^2) T(k_\perp^2,\mu^2)\right],
\end{equation}
which then corresponds to the probability of finding a gluon in the
proton with transverse momentum $k_\perp$ and energy fraction $x$ when
probed with a hard scale $\mu^2$. $T$ is the survival probability
of the gluon given by the Sudakov form factor,
\begin{equation}
  \label{eq:mishasudakov}
  \ln T(k_\perp^2,\mu^2) =
  -\int_{k_\perp^2}^{\mu^2}\frac{dq_\perp^2}{q_\perp^2}
  \frac{\alpha_S(q_\perp^2)}{2\pi}\int_{0}^{\frac{\mu}{\mu+q_\perp}}
  dz\left[zP_g(z) + n_f P_q(z)\right].
\end{equation}
To get the oduPDF one then starts from \eqref{eq:pdf2odpdf} and get by
analogy in the limit $x'\ll x$
\begin{equation}
  \label{eq:odpdf2odupdf}
  f_g(x,x',k_\perp^2,\mu^2)\approx
  \frac{d}{d\ln k_\perp^2}\left[R_g xg(x,k_\perp^2)
    \sqrt{T(k_\perp^2,\mu^2)}\right],
\end{equation}
where the square root of the Sudakov comes about because only one of
the two gluons are probed by the hard scale.

The hard scale $\mu$ in the oduPDF and in the Sudakov form factor is
in the KhMR approach argued to be $m_H/2$. In fact the number is
$0.62\cdot m_H$ and comes from a tuning to reproduce full one-loop
vertex corrections \cite{Kaidalov:2003fw}. For LDC, below, we will be
less ambitious and simply use $m_H$ as scale.

\section{Unintegrated parton densities}
\label{sec:unint-part-dens}

A general comment concerning the unintegrated gluon densities used in
the KhMR calculations is that the KMR prescription essentially
corresponds to taking one step backward in a DGLAP-based initial-state
parton shower, to \textit{unintegrate} the integrated PDF. As
mentioned in the introduction, there are indications that such a
prescription underestimates the hardness of the \kT-distribution of
the uPDF. Therefore we will here investigate uPDFs based on CCFM and
LDC evolution, where emissions unordered in \kT\ are allowed, which
could increase the hardness of the \kT-distribution.

\subsection{The Linked Dipole Chain uPDF}
\label{sec:LDC}

The Linked Dipole Chain model\cite{Andersson:1996ju,Andersson:1998bx}
is a reformulation and generalization of the
CCFM\cite{Ciafaloni:1988ur,Catani:1990yc,Catani:1990sg,Marchesini:1995wr}
evolution for the unintegrated gluon.  CCFM has the property that it
reproduces BFKL evolution\cite{Kuraev:1977fs,Balitsky:1978ic} for
asymptotically large energies (small $x$) and is also similar to
standard DGLAP evolution
\cite{Gribov:1972ri,Lipatov:1975qm,Altarelli:1977zs,Dokshitzer:1977sg}
for larger virtualities and larger $x$. It does this by carefully
considering coherence effects between gluons emitted from the
evolution process, allowing only gluons ordered in angle to be emitted
in the initial state, and thus contribute to the uPDFs, while
non-ordered gluons are treated as final state radiation off the
initial state gluons. LDC differs from CCFM by the fact that it is
ordered both in positive and negative light cone momenta, $q_+$ and
$q_-$, of the emitted gluons, a treatment which categorizes more
emissions as final state emission as compared to CCFM. This symmetric
ordering in both $q_+$ and $q_-$, which also implies ordering in
rapidity $y$ or angle, together with the additional requirement that
the transverse momentum of an emitted gluon must be larger than the
\kT\ of the propagator gluon before or after the emission, greatly
simplifies the evolution equations and has as a consequence that the
uPDF approximately factorizes into a one-scale density multiplied by
the Sudakov form factor:
\begin{equation}
  \label{eq:fact}
  G(x,\kT^2,\mu^2)\approx G(x,\kT^2)\times\Delta_S(\kT^2,\mu^2),
\end{equation}
where
\begin{equation}
  \label{eq:sudLDC}
  \ln\Delta_S(k_\perp^2,M^2)=
  -\int_{k_\perp^2}^{M^2}\frac{dq_\perp^2}{q_\perp^2}
  \frac{\alpha_s}{2\pi}
  \int_0^{1-q_\perp/M}dz\left[zP_g(z) + \sum_qP_q(z)\right].
\end{equation}

The LDC model has been implemented in an event generator which is then
able to generate complete events in DIS with final state radiation
added according to the dipole cascade model
\cite{Gustafson:1986db,Gustafson:1988rq} and hadronization according
to the Lund model \cite{Andersson:1983ia}. One advantage of having an
event generator implementation is that energy and momentum can be
conserved in each emission. Since the lack of momentum conservation in
the BFKL formalism is the main reason for the huge next-to-leading
logarithmic corrections\cite{Fadin:1998py}, the LDC model is therefore
expected to have smaller sub-leading corrections (see
\cite{Andersson:2002cf} for a more detailed discussion on this).

The perturbative form of the uPDF needs to be convoluted with
non-perturbative input PDFs, the form of which are fitted to reproduce
the experimental data on $F_2$. This has all been implemented in the
\ldcmc program \cite{Kharraziha:1998dn,Kharraziha:ldcmc}, and the
resulting events can be compared directly to experimental data from
eg.\ HERA.  The LDC gluon uPDF can then be extracted by generating DIS
events with \ldcmc and measuring the gluon density as described in
\cite{Gustafson:2002jy}.  Due to the \kT-unordered nature of the LDC
evolution, the relationship between the uPDF and the standard gluon
density is different from \eqref{eq:pdf2updf}, as the integrated gluon
at a scale $\mu^2$ also receives a contribution, although suppressed,
from gluons with $k_\perp>\mu$, and in \cite{Gustafson:2002jy} the
following expression was obtained:
\begin{eqnarray}
  \label{eq:updf2pdf}
  xg(x,\mu^2)&=&G_0(x)\Delta_S(k_{\perp0}^2,\mu^2)\\
  &+&\int_{k_{\perp0}^2}^{\mu^2}\frac{dk_\perp^2}{k_\perp^2}
  G(x,k_\perp^2)\Delta_S(k_\perp^2,\mu^2) +
  \int_{\mu^2}^{\mu^2/x}\frac{dk_\perp^2}{k_\perp^2}
  G(x\frac{k_\perp^2}{\mu^2},k_\perp^2)\frac{\mu^2}{k_\perp^2},\nonumber
\end{eqnarray}
where $G_0(x)$ is the non-perturbative input parameterization at the
cutoff scale \kTi{0}.

Note that a sharp cutoff \kTi{0}\ is assumed, which could cause
problems in calculations sensitive to the small-\kT\ behavior. To avoid
this we redefine the uPDF as
\begin{equation}
  \label{eq:LDCuPDF}
  G(x,\kT^2,\mu^2)=\left\{\begin{array}{ll}
        a\left(\frac{\kT^2}{\kTi{0}^2}\right)^a
        G_0(x)\Delta_S(k_{\perp0}^2,\mu^2) & \kT < \kTi{0} \\[3mm]
        G(x,k_\perp^2)\Delta_S(k_\perp^2,\mu^2) & \kTi{0}<\kT<\mu\\[3mm]
        G(x\frac{k_\perp^2}{\mu^2},k_\perp^2)\frac{\mu^2}{k_\perp^2} &
        \mu<\kT<\mu/\sqrt x
      \end{array}\right.,
\end{equation}
where $a$ can either be set to $1$, as was effectively done in
\cite{Lonnblad:2003wx}, or to $G(x,\kTi{0}^2)/G_0(x)$ which
makes the distribution continuous across \kTi{0}. In this way we get
the standard form
\begin{equation}
  \label{eq:updf2pdf2}
  xg(x,\mu^2)=\int_0^\infty\frac{dk_\perp^2}{k_\perp^2} G(x,k_\perp^2,\mu^2),
\end{equation}
and we find that our results are not very sensitive to the choice of
$a$.

To obtain the off-diagonal densities needed for the exclusive
luminosity function, we assume that a similar approximation can be
made as for the KMR densities, that is, in the limit of very small $x'$
\begin{equation}
  \label{eq:oduLDC}
  f_g\sup{LDC}(x, x', \kT^2, \mu^2)\approx
  R_g G(x,\kT^2)\sqrt{\Delta_S(\kT^2,\mu^2)}.
\end{equation}
The square root of the Sudakov form factor is used, since only one of
the gluons couples to the produced Higgs at the high scale, and we
could equivalently have written
\begin{equation}
  \label{eq:oduLDC2}
  f_g\sup{LDC}(x, x', \kT^2, \mu^2)\approx
  R_g \sqrt{G(x,\kT^2,\mu^2)G(x,\kT^2,\kT^2)}.
\end{equation}
We note that this is not completely equivalent to
\eqref{eq:odpdf2odupdf}, but it is a prescription which can be used
for any uPDF, not only the KMR one. Using \eqref{eq:oduLDC2} rather
than \eqref{eq:odpdf2odupdf} for the KMR uPDFs we find that the
exclusive luminosity function is underestimated by $\approx50\%$ for a
higgs mass of 120~GeV.

\subsection{The Jung 2003 uPDF parameterizations}
\label{sec:Jung}

The Jung 2003\cite{Hansson:2003xz} unintegrated parton distribution
functions are based on standard CCFM evolution and was obtained using
a Monte Carlo implementing forward evolution\footnote{Based on the
  \smallx program \cite{Marchesini:1990zy,Marchesini:1992jw}.}. The
main difference w.r.t.\ LDC is, as mentioned above, that CCFM allows
more emissions in the initial state, which makes it more infrared
sensitive and which prevents the simple factorization into a one-scale
density and a Sudakov form factor as in \eqref{eq:fact}. Another
difference is that CCFM only describes gluon evolution, while in the
LDC it is also possible to include quarks.

Just as for LDC, the perturbative CCFM evolution needs to be convoluted
with non-perturbative input parton density, the parameters of which
are determined by a fit to \ftwo at small $x$ determined at HERA.

To produce full events the Jung uPDFs may be convoluted with an
appropriate off-shell matrix element (eg.\ $\gamma^*g^*\rightarrow
\q\qbar$) and the final state partons can then be generated in a
backward evolution algorithm implemented in the \cascade
program\cite{Jung:2001hx}.

To obtain the off-diagonal densities, we use the same procedure as in
LDC given by \eqref{eq:oduLDC2},
\begin{equation}
  \label{eq:oduJung}
  f_g\sup{Jung}(x, x', \kT^2, \mu^2)\approx
  R_g \sqrt{G(x, \kT^2, \mu^2)G(x, \kT^2, \kT^2)},
\end{equation}
but note that the equivalence with \eqref{eq:oduLDC} does not hold
since the factorization in \eqref{eq:fact} is absent in the Jung
uPDFs.

\subsection{Summary of uPDFs}
\label{sec:summary-updfs}

Within the three different procedures for obtaining uPDFs there are a
number of optional behaviors to choose from which are summarized in
table \ref{tab:updfs}. For KMR we can choose different integrated
densities to start from, but that has already been shown to only give
rise to moderate differences\cite{Khoze:2001xm}.  Since the integrated
PDFs have been fitted to a wide range of inclusive data, the
description of such observables are trivially also reproduced by the
KMR uPDFs. For less inclusive observables the situation is less clear,
and as argued in the introduction there are indications that the KMR
procedure will underestimate slightly the hardness of the
\kT-distribution especially at small \kT. And although it has been
showed to be able to reproduce inclusive jet cross sections in deeply
inelastic scattering at HERA \cite{Watt:2003mx}, it is not likely that
it will be able to explain the forward jet rates with $\kTi{jet}^2\sim
Q^2$.

In \cite{Lonnblad:2003wx} we used the three different options for the
LDC densities introduced in \cite{Gustafson:2002jy}, which differ in
the splitting functions included in the evolution. The \standard
option includes all splitting functions and hence includes also the
evolution of quarks. The \gluonic and \leading options only includes
gluons and differs in that the latter only includes the leading $1/z$
and $1/(1-z)$ terms in the gluon splitting function. All give
reasonable fits to HERA \ftwo measurements in the region $x<0.01$ and
1~GeV$^2\ltaeq Q^2\ltaeq100$~GeV$^2$. The \standard option also
describes \ftwo at higher $x$ values where the contribution of valence
quarks is more important. Of the three only the \leading is able to
satisfactorily describe forward jets indicating that the other two
probably underestimates the hardness of the \kT-distribution of the
gluon.

The Jung 2003 distributions also come with different options. Here we
will use Jung-1 and Jung-2 which are similar to the LDC \leading and
\gluonic options respectively in that the former only uses the leading
terms in the gluon splitting functions, while the latter uses the full
splitting function. Also these give a good description of \ftwo in the
fitted region of $x<0.01$ and 1~GeV$^2\ltaeq Q^2\ltaeq100$~GeV$^2$.
When used in the \cascade generator, only the Jung-1 is able to give a
good description of forward jets. Also for other observables the Jung
uPDFs give results which are consistent with the ones obtained with
LDC.

\TABLE[t]{
  \begin{tabular}{|l|l|l|l|l|}
    \hline
    uPDF & evolution & splittings & inclusive observables & forward jets \\
    \hline
    \hline
    KMR & DGLAP & full & globally good fit & probably not \\
    \hline
    \standard & LDC & full & HERA \ftwo & no \\
    \hline
    \gluonic & LDC & full gluon & HERA \ftwo at small $x$ & no \\
    \hline
    \leading & LDC & singular gluon & HERA \ftwo at small $x$ & yes \\
    \hline
    Jung-1 & CCFM & singular gluon & HERA \ftwo at small $x$ & yes \\
    \hline
    Jung-2 & CCFM & full gluon & HERA \ftwo at small $x$ & no \\
    \hline
  \end{tabular}\caption{\label{tab:updfs} Summary of the different uPDFs
    used in this report, indicating the differences in evolution and
    the ability to reproduce experimental observables.}  }

\section{Results}
\label{sec:results}

Armed with these six uPDFs and their corresponding off-diagonal
densities, we now want to see how they influence the exclusive
luminosity function at LHC energies. But before we do this we want to
compare the uPDFs in general to see if they at all make sense at the
scales involved when considering Higgs production at LHC.

\subsection{Inclusive Higgs production}
\label{sec:incl-higgs-prod}

\FIGURE[t]{\epsfig{file=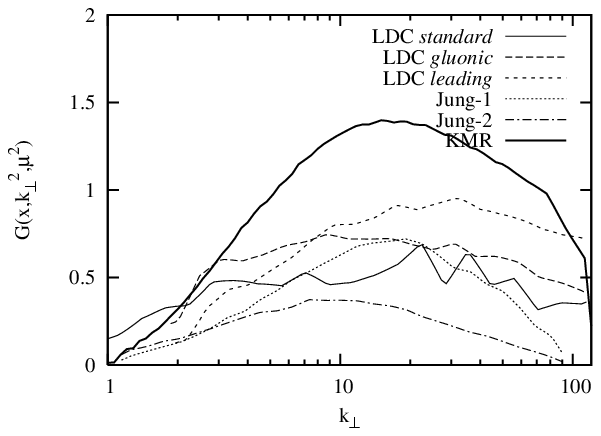,width=10cm}
  \caption{\label{fig:G120kt} The unintegrated gluon densities as a function
    of \kT\ for $\mu=m_H=120$~GeV and
    $x=x_H=m_H/\sqrt{S}\approx0.086$.  The full line is LDC \standard,
    long-dashed is LDC \gluonic, short-dashed LDC \leading, dotted is
    Jung-1, dash-dotted is Jung-2 and the thick full line is KMR. The
    wiggly shape of the LDC curves is due to low statistics when
    extracting them in \cite{Gustafson:2002jy}.}}

First we look in figure \ref{fig:G120kt} at the uPDFs relevant for
producing a central exclusive 120~GeV Higgs at the LHC, i.e.\ 
$x=x_H=m_H/\sqrt{S}$ and $\mu=m_H=120$~GeV. What is shown is the
logarithmic density in \kT\ and clearly there are large differences
between the uPDFs both in shape and normalization. For the shape the
LDC densities stick out as they do not tend to zero for
$\kT\rightarrow\mu$. This is as expected for LDC evolution with
unordered \kT-evolution. CCFM will also allow $\kT>\mu$, but it seems
that this is more suppressed for high scales.  For the shapes we can
also imagine a rough agreement between \standard, \gluonic and Jung-2,
while \leading and Jung-1 are clearly harder. This is also expected as
the absence of nonsingular terms in \leading and Jung-1 enhances the
radiation from gluons.

\FIGURE[t]{
  \epsfig{file=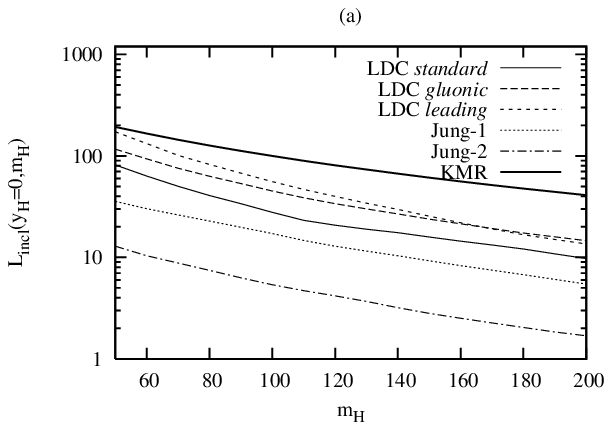,width=7.5cm}\epsfig{file=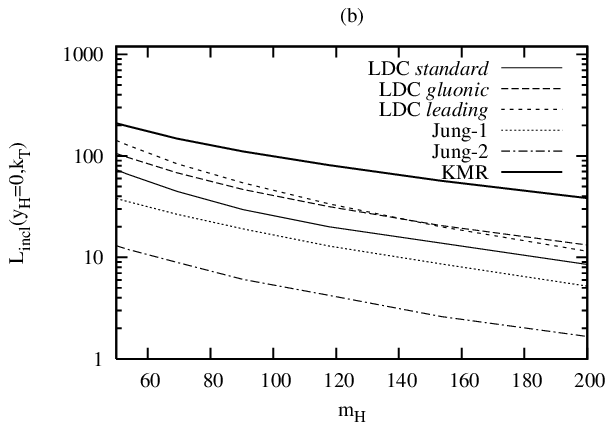,width=7.5cm}
  \caption{\label{fig:LinmH} The inclusive gluon luminosity for
    central Higgs production as a function of $m_H$. (a) is simply the
    square of the integrated gluon density, while (b) is properly
    integrated over \kT\ and includes the \kT-dependence of the
    off-shell matrix element. The lines are the same as in figure
    \ref{fig:G120kt}}}

The difference in normalization also shows up in the predictions for
inclusive Higgs production. This is shown in figure \ref{fig:LinmH}.
In figure \ref{fig:LinmH}a we show the square of the integrated gluon
densities, which would enter in a calculation using collinear
factorization. In figure \ref{fig:LinmH}b we use the \kT-dependence of the
off-shell matrix element given in \cite{Hautmann:2002tu}:
\begin{equation}
  \label{eq:offshellH}
  \hat{\sigma}^*(m_H, \vec{k}_{\perp1}, \vec{k}_{\perp2})=
  \hat{\sigma}_0\cdot 2
  \left(\frac{m_{\perp H}^2\cos(\phi)}{m_H^2+\kTi{1}^2+\kTi{2}^2}\right)^2,
\end{equation}
where $\vec{k}_{\perp1}$ and $\vec{k}_{\perp2}$ are the incoming
transverse momenta, $\phi$ the angle between them, $m_{\perp H}$ the
resulting transverse mass of the Higgs, and $\hat{\sigma}_0$ is the
standard on-shell matrix element. This gives us the inclusive
luminosity function,
\begin{equation}
  \label{eq:inlumi}
  L(m_H,y)=\int\frac{d\kTi{1}^2}{\kTi{1}^2}\frac{d\kTi{2}^2}{\kTi{2}^2}d\phi
  \frac{\hat{\sigma}^*}{\hat{\sigma}_0}
  G(x_1,\kTi{1}^2,\mu^2)G(x_2,\kTi{2}^2,\mu^2),
\end{equation}
where $x_{1,2}=m_{\perp H}e^{\pm y}$ and $\mu=m_{\perp H}$. We use
this scale also for KMR, since this is what was used in the case of W
and Z production \cite{Watt:2003vf}. As seen in figure \ref{fig:LinmH}
there are small, but not insignificant differences between the
collinear and off-shell versions. In fact the off-shell version takes
into account some of the beyond leading order effects which are absent
in our LO collinear approximation.

Clearly the differences in the inclusive luminosity are too large to
be taken as genuine uncertainties in the prediction for the Higgs
cross section. For such integrated quantities we expect the standard
DGLAP approach implemented in KMR to give a reasonably predictive
answer, and we conclude that the CCFM and LDC based densities
parameterizations simply are not well enough constrained to give
reasonable predictions for Higgs production at the LHC. The problem is
that the Jung and LDC densities have only been fitted to \ftwo at HERA
which means mainly small $x$ and \qsq, while for Higgs production we
have much larger scales and through evolution we are also sensitive to
the large-$x$ behavior at lower scales, which is not well constrained.

If the LDC and Jung densities are not constrained enough to predict
inclusive Higgs production at the LHC, it is unlikely that they are
able to say anything predictive about exclusive Higgs production.
However, although the normalization is uncertain, it may still be
possible that these densities have some predictive power on the
\kT-dependence of the uPDF. In figure \ref{fig:ktH} we show the
normalized \kT-distribution of a centrally produced Higgs at the LHC
as predicted by the different uPDFs, and we see that the differences
are large, but not unreasonable. We find that the spectra are harder
for \leading and Jung-1 than for \standard, \gluonic and Jung-2, which
is expected since the former only have singular terms in the gluon
splitting function which allows the gluon to radiate more.

\FIGURE[t]{\epsfig{file=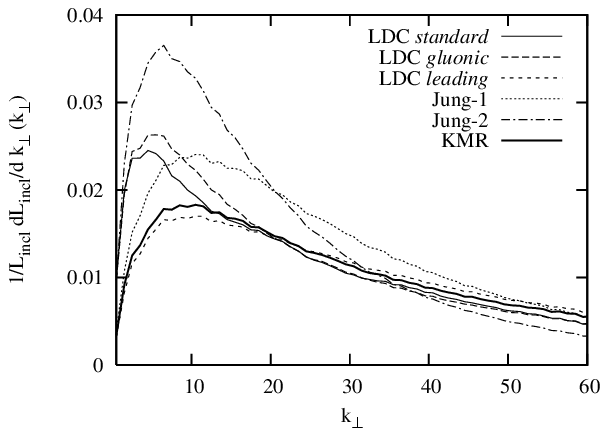,width=10cm}
  \caption{\label{fig:ktH} The normalized \kT-distribution
    of a central Higgs produced at LHC as predicted by using different
    uPDFs. The lines are the same as in figure \ref{fig:G120kt}.}}

\subsection{Exclusive Higgs production}
\label{sec:excl-higgs-prod}

Although we do not believe that the LDC/Jung uPDFs can be used to give
any prediction for neither the inclusive or exclusive luminosity, it
is not unlikely that they actually have some predictive power on the
ratio of the two. We saw above that the normalized \kT-distribution of
the Higgs looks reasonable. In addition, although the uPDFs enters to
the power 4 in the exclusive luminosity function, according to
\eqsref{eq:oduLDC2} and (\ref{eq:oduJung}), the high scale uPDF only
enters with power 2 while the other two powers depend on lower scales
where the uPDFs may be better constrained. Hence, the uncertainty from
the evolution to high scales may cancel in the ratio.

\FIGURE[t]{
  \epsfig{file=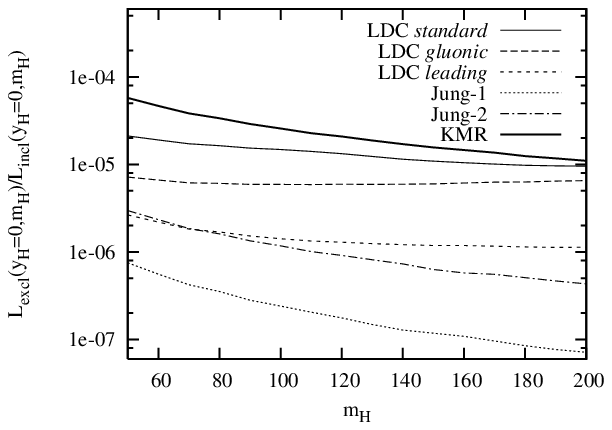,width=10cm}
  \caption{\label{fig:LrmH} The ratio of the exclusive to inclusive
    luminosity function at fixed central rapidity as a function of
    $m_H$.  The lines are the same as in figure \ref{fig:G120kt}.}}

In figure \ref{fig:LrmH} we show the ratio between the exclusive and
the inclusive luminosity functions for fixed central rapidity as a
function of $m_H$ according to \eqsref{eq:kmrlum} and
(\ref{eq:inlumi}). There are clearly large differences, probably too
large to be attributed to anything else than that the LDC and Jung
densities simply are not constrained enough to give any reasonable
predictions.

We know that the inclusive luminosity in \eqref{eq:kmrlum} is mostly
sensitive to \kT-values around a couple of GeV, and we can see that
the Jung-1 is much lower than Jung-2 which can be attributed to the
fact that Jung-1 has a harder \kT-distribution than Jung-2 reducing
the density in this region relative to higher \kT. Similarly \leading
is much lower than \standard and \gluonic and again the former has a
harder \kT-distribution than the two latter. But since there are large
differences in general between LDC and Jung we cannot say that the
differences simply does not come from the fact that all these uPDFs
are too unconstrained.

To focus on the uncertainties in the \kT\ distribution of the uPDFs, we
instead concentrate only on the KMR densities, where we know that the
overall normalization is well constrained, and study what happens if we
simply shift the \kT\ distribution slightly, while keeping the
integrated PDF fixed. We know that the \kT-spectrum of the Z and W at
the Tevatron can be well described by standard DGLAP based parton
showers if an Gaussian intrinsic \kT\ with a width of a couple of GeV
is added to the incoming quarks. Judging from figure \ref{fig:watt} it
does not seem unlikely that the shape would be better reproduced if
the KMR uPDF was modified in the same way.

\FIGURE[t]{\epsfig{
    file=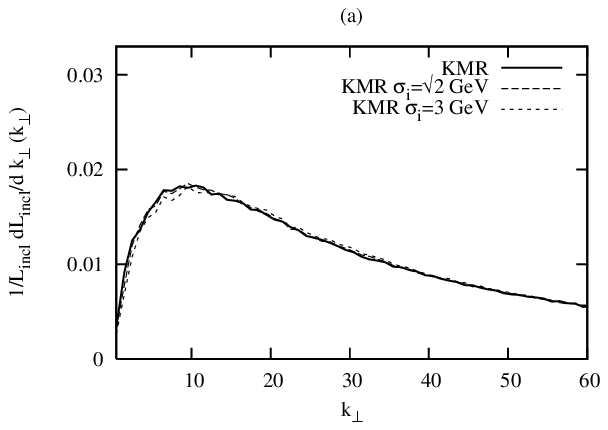,width=7.5cm}\epsfig{file=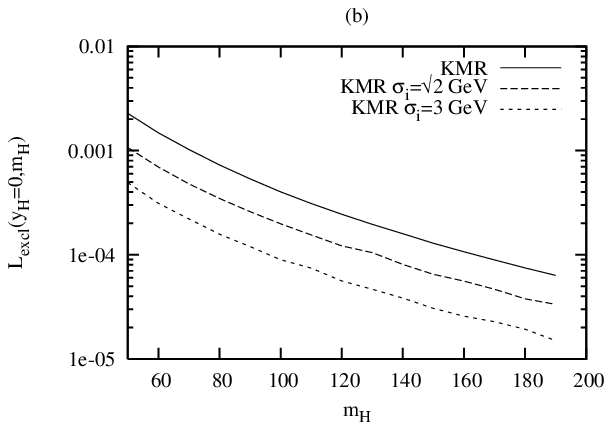,width=7.5cm}
  \caption{\label{fig:LexmH} (a) The normalized \kT-distribution
    of a central Higgs produced at LHC as predicted by using KMR uPDFs
    with different Gaussian intrinsic \kT\ added. (b) The exclusive
    luminosity function (\eqref{eq:kmrlum}) at fixed central rapidity
    as a function of $m_H$ calculatedusing KMR uPDFs with different
    Gaussian intrinsic \kT\ added. In both cases the full line
    corresponds to no intrinsic \kT\ and dashed and dotted lines
    corresponds to a Gaussian intrinsic \kT\ with a width of
    $\sqrt{2}$ and 3~GeV respectively.}}

In figure \ref{fig:LexmH}a we see the effect of such an intrinsic \kT\ 
on the \kT-distribution of a 120~GeV Higgs at fixed central rapidity
at LHC. We here use a larger intrinsic \kT\ than would be needed at
the Tevatron which, as discussed above, is not unreasonable since we
are here dealing with gluons rather than quarks and we have much
smaller $x$-values, allowing for more unordered evolution. Still the
effect on the Higgs spectrum is rather moderate, especially compared
to the effects in figure \ref{fig:ktH}.  However, the effect on the
exclusive luminosity is large, as can be seen in figure
\ref{fig:LexmH}b. Adding a Gaussian intrinsic \kT\ with a width of
3~GeV reduces the luminosity by approximately a factor 5.  And we
conclude that the exclusive production of Higgs at the LHC is very
sensitive to the small-\kT\ distribution of the unintegrated gluon.


\section{Conclusions}
\label{sec:conclusions}

The main conclusion of this article is a negative one. The predictive
powers of the unintegrated gluon density functions as fitted only to
small-$x$ HERA data is very poor when applied to exclusive Higgs
production at LHC. In fact, not even inclusive Higgs production at the
LHC is well constrained with these uPDFs. However, looking at the
qualitative differences between these uPDFs we can learn something
about where the uncertainties come from. Here we have argued that
there are problems not only with the overall normalization of the
uPDFs at the high scales under consideration, but also the actual
\kT-distribution at small \kT\ is important. The reason is clearly
visible in the \kT-integration in the exclusive luminosity function,
where the main contribution comes from transverse momenta in the
region of a couple of GeV.

The situation is quite different when it comes to the uPDF derived
from the integrated gluon density using the KMR prescription. Here we
believe the overall normalization to be well determined by the global
PDF fits, and the predictions for inclusive Higgs production should be
trustworthy. However, the prediction for the distribution of small \kT
values is less certain and there is evidence that eg.\ the \kT
distribution for W and Z production at the Tevatron obtained from the
KMR prescription is a bit to high for small \kT. This is consistent
with the behavior of DGLAP-based parton shower approaches, which are
closely related to the KMR approach, which typically need an
additional gaussian intrinsic \kT\ of one or two GeV to reproduce W and
Z transverse momentum spectra. We have found that introducing an
intrinsic \kT\ in the KMR uPDF in the calculation of the exclusive
luminosity function will give a clear reduction.


We will not try to use our findings to make an estimate of the
uncertainties involved in the KhMR predictions for the exclusive Higgs
production at the LHC and elsewhere. Clearly there is a need to find
better experimental observables to constrain the (off-diagonal)
unintegrated gluon density before we can make precise predictions. We
do feel that the published KhMR predictions may be too high, but
clearly they should give the right order of magnitude, and the
prospect of using the exclusive process to study the Higgs at the LHC
is still a very interesting one.

\section*{Acknowledgments}

We would like to thank Hannes Jung, Valery Khoze and Misha Ryskin for
valuable discussions.

\bibliographystyle{utcaps}  
\bibliography{references} 

\end{document}